\documentstyle[12pt]{article}

\input epsf

\begin{document}

\begin{flushright}
 CECS-PHY-99/22 \\
hep-th/9912259 
\end{flushright}
\vspace{0.4cm}

\begin{center}
{\large {\bf Charged Rotating Black Hole in Three Spacetime Dimensions}}

\vspace{1.0cm}

{\sc Cristi\'{a}n Mart\'{\i }nez$^{\,a}$, Claudio Teitelboim$^{\,a,b}$\\[0pt]
and Jorge Zanelli$^{\,a,c}$ }\\[0pt]

\vspace{0.9cm}

{\em $^{a}$Centro de Estudios Cient\'{\i}ficos de Santiago,} {\em Casilla
16443, Santiago 9, Chile } \\[10pt]
{\em $^{b}$Institute for Advanced Study,} {\em Olden Lane, Princeton, New
Jersey 08540, USA} \\[10pt]
{\em $^{c}$Universidad de Santiago de Chile, Casilla 307, Santiago 2, Chile }
\\[0pt]

\vspace{1.0cm}

\centerline{\bf Abstract} \vspace{- 4 mm}
\end{center}

\begin{quote}
The generalization of the black hole in three-dimensional spacetime to
include an electric charge $Q$ in addition to the mass $M$ and the angular
momentum $J$ is given. The field equations are first solved explicitly when $Q$ 
is small and the general form of the field at large distances is
established. The total ``hairs'' $M$, $J$ and $Q$ are exhibited as boundary
terms at infinity. It is found that the inner horizon of the rotating
uncharged black hole is unstable under the addition of a small electric
charge. Next it is shown that when $Q=0$ the spinning black hole may be
obtained from the one with $J=0$ by a Lorentz boost in the $\varphi -t$
plane. This boost is an ``illegitimate coordinate transformation'' because it
changes the physical parameters of the solution. The extreme black hole
appears as the analog of a particle moving with the speed of light. The same
boost may be used when $Q\neq 0$ to generate a solution with angular
momentum from that with $J=0$, although the geometrical meaning of the
transformation is much less transparent since in the charged case the black holes
are not obtained by identifying points in anti-de Sitter space. The metric
is given explicitly in terms of three parameters, $\widetilde{M}$, $%
\widetilde{Q}$ and $\omega $ which are the ``rest mass'' and ``rest charge''
and the angular velocity of the boost. These parameters are related to $M$, $%
J$ and $Q$ through the solution of an algebraic cubic equation.
Altogether, even without angular momentum, the electrically charged 2+1
black hole is somewhat pathological since (i) it exists for{\bf \ }%
arbitrarily negative values of the mass, and (ii) there is no upper bound on
the electric charge.

\vfill
\end{quote}


\section{Introduction}

\label{introduction}

The black hole in three-dimensional spacetime \cite{BTZ,BHTZ} has been the
object of considerable study \cite{Carlip}. However --as pointed out by
several authors \cite{Carlip}, \cite{Others}-- if one includes electric
charge, the solution reported in the original article \cite{BTZ} only
applies when the angular momentum vanishes. In this article we analyze the
case when all three ``hairs'' $M$, $J$ and $Q$ are different from zero.

The work is organized as follows. Section 2 presents the Einstein-Maxwell
action in hamiltonian form. The hamiltonian form is well suited for the
present analysis because it permits to identify directly the physical
meaning of the integration constants as the total charges and their
conjugates. The requirements of circular symmetry and time independence
(stationarity) are then imposed and the simplified equations applying in
that case are written down. The equations are a coupled set of first order
differential equations. If the electric charge is small the equations can be
integrated directly. This is done in Sec. 3. A noteworthy fact is that the
inner horizon of the rotating uncharged black hole is unstable under the
addition of a small electric charge.

Section 4 begins by showing that when $Q=0$ the spinning black hole may be
obtained from the one with $J=0$ by a Lorentz boost in the $\varphi -t$
plane. This boost is an ``illegitimate coordinate transformation '' because it
changes the physical parameters of the solution. The extreme black hole
appears as the analog of a particle moving with the speed of light. The same
boost may be used when $Q\neq 0$ to generate a solution with angular
momentum from that with $J=0$, although the geometrical meaning of the
transformation is much less transparent since in that case the black holes
are not obtained by identifying points in anti-de Sitter space. The metric
is given explicitly in terms of three parameters, $\widetilde{M}$, $\widetilde{Q}$
 and $\omega $ which are the ``rest mass'' and ``rest charge''
and the angular velocity of the boost. These parameters are related to
 $M$, $J$ and $Q$ through the solution of an algebraic cubic equation.
Altogether, even without angular momentum, the electrically charged 2+1
black hole is somewhat pathological since ({\bf i}) it exists for
arbitrarily negative values of the mass, and ({\bf ii}) there is no upper
bound on the electric charge\footnote{ As it was observed
 in \cite{Coussaert-Henneaux}, the lack of a
lower bound for the mass prevents charged black holes from being
supersymmetric.}.

\section{Einstein-Maxwell equations}

\label{E-Meqs}


\subsection{Hamiltonian form}

The Hilbert action coupled to electromagnetism
\begin{equation}
I_{{\rm H}}=\int d^{3}x\sqrt{-g}\left( \frac{R-2\Lambda }{2\kappa }-\frac{1}{%
4}F_{\mu \nu }F^{\mu \nu }\right) ,  \label{Lagrangian}
\end{equation}
can be cast in hamiltonian form,
\begin{equation}
I=\int dtd^{2}x\left( \pi ^{ij}\dot{g_{ij}}+{\cal E}^{i}\dot{A}_{i}-N^{\perp}%
{\cal H}_{\perp }-N^{i}{\cal H}_{i}-A_{0}G\right) \,,  \label{hamilaction}
\end{equation}
by adding appropriate surface terms.

The canonical variables are the spatial metric $g_{ij}$ and the vector
potential $A_{i}$ together with their conjugate momenta $\pi ^{ij}$ and $%
{\cal E}^{i}$. The generators of normal deformations (${\cal H}_{\perp }$),
tangential deformations (${\cal H}_{i}$) and gauge transformations ($G$) are
given by
\begin{eqnarray}
{\cal H}_{\perp } &=&2\kappa g^{-1/2}(\pi ^{ij}\pi _{ij}-(\pi
_{i}^{i})^{2})-(2\kappa )^{-1}g^{1/2}(R-2\Lambda )\   \nonumber
\label{hperp} \\
&+&\frac{1}{2}g^{-1/2}(g_{ij}{\cal E}^{i}{\cal E}^{j}+{\cal B}^{2}) \\
{\cal H}_{i} &=&-2{\pi _{i}^{j}}_{|j}-\epsilon _{ji}{\cal E}^{j}{\cal B}
\label{hi} \\
G &=&-{\cal E}_{\,,i}^{i}\,.  \label{GG}
\end{eqnarray}
In terms of the lapse ($N^{\perp }$) and shift ($N^{i}$) Lagrange
multipliers and the spatial two-dimensional metric $g_{ij}$, the space-time
line element is
\begin{equation}
ds^{2}=-(N^{\perp })^{2}dt^{2}+g_{ij}(dx^{i}+N^{i}dt)(dx^{j}+N^{j}dt)\,,
\end{equation}
and the magnetic density ${\cal B}$ is given by
\begin{equation}
F_{ij}=\epsilon _{ij}{\cal B}\,.
\end{equation}
We will use units such that the gravitational constant $\kappa =8\pi G$,
which has dimensions of an inverse mass, is set equal to $\frac{1}{2}$. The
cosmological constant is negative and is related to the cosmological length $%
l$ by $\Lambda =-l^{-2}$. This length will be set equal to unity.


\subsection{Rotational symmetry and time independence}

\label{circularS}

If one requires invariance under spatial rotations and time translations,
the metric may be cast in the form
\begin{eqnarray}
ds^{2} &=&-N^{2}(r)f^{2}(r)dt^{2}+f^{-2}(r)dr^{2}+r^{2}(d\varphi +N^{\varphi
}(r)dt)^{2},  \nonumber  \label{axial} \\
&&0\leq r<\infty ,\;\;\;\;0\leq \varphi <2\pi ,\;\;\;\;t_{1}\leq t\leq
t_{2}\,.
\end{eqnarray}
This implies that the only non-vanishing component of the gravitational
momentum is given by
\begin{equation}
\pi {_{\varphi }}{^{r}}=p(r)\,.  \label{momentum}
\end{equation}
The only non-vanishing electromagnetic strengths are
\begin{equation}
{\cal E}^{r}=Q\,,  \label{Er}
\end{equation}
the total electric charge, and
\begin{equation}
{\cal B}=\partial _{r}A_{\varphi } \,.  \label{Bs}
\end{equation}
Equation (\ref{Er}) follows from solving the constraint $G=0$, whereas $%
{\cal E}^{\varphi }=0$ follows from ${\cal H}_{r}=0$. To arrive at (\ref{Bs}%
) one uses the radial gauge $A_{r}=0$.

In terms of the remaining fields, which are only functions of $r$, the
action reads
\begin{equation}
I=-(t_{2}-t_{1})2\pi \int \left( N{\cal H}+N^{\varphi }{\cal H}_{\varphi
}\right) dr\,,  \label{redhamilaction}
\end{equation}
with
\begin{equation}
{\cal H}=f{\cal H}_{\perp }=(f^{2})^{\prime }-2r+\frac{2p^{2}}{r^{3}}+ \frac{%
Q^{2}}{2r}+\frac{f^{2}{\cal B}^{2}}{2r} \,,  \label{Hone}
\end{equation}
and
\begin{equation}
{\cal H}_{\varphi }=-2p^{\prime }-Q{\cal B}\,.  \label{Htwo}
\end{equation}
Extremization of the reduced action (\ref{redhamilaction}) with respect to $%
N $, $N^{\varphi }$, $f^{2}$, $p$ and $A_{\varphi }$ gives, respectively,
\begin{eqnarray}
(f^{2})^{\prime }-2r+\frac{2p^{2}}{r^{3}}+\frac{Q^{2}}{2r}+\frac{f^{2}{\cal B%
}^{2}}{2r} &=&0  \label{one} \\
2p^{\prime }+Q{\cal B} &=&0  \label{two} \\
N^{\prime }-\frac{N{\cal B}^{2}}{2r} &=&0  \label{three} \\
(N^{\varphi })^{\prime }+\frac{2Np}{r^{3}} &=&0  \label{four} \\
\left( \frac{Nf^{2}{\cal B}}{r}-N^{\varphi }Q\right) ^{\prime } &=&0.
\label{te1}
\end{eqnarray}


\section{Solutions in special cases and physical \\ meaning of integration
constants}

Equation (\ref{te1}) can be solved for ${\cal B}$ with an appropriate
integration constant which is fixed by regularity and is not an independent
free parameter [see Eq. (\ref{te4}) below]. The integration constants for $%
f^{2}$ and $p$ can be expressed in terms of the mass $M$ and the angular
momentum $J$. Thus, together with the electromagnetic field, the solution is
characterized by the charges $M$, $J$, $Q$ and their conjugates $N(\infty )$%
, $N^{\varphi }(\infty )$ and $A_{0}(\infty )$ (``chemical potentials'' in
the euclidean formulation).

The system of first order nonlinear differential equations (\ref{one}-\ref
{four}) is not straightforward to solve for $f^{2}$, $p$, $N$ and $%
N^{\varphi}$, when $Q\neq 0$. It is therefore useful to analyze special
cases and approximations in order to gain insight into the properties of the
solution.

We first turn our attention to the two limiting cases when the solution is
known in closed form. These are $Q=0$, $J\neq 0$, $M\neq 0$ (the standard
2+1 black hole) and $Q\neq 0$, $J=0$, $M\neq 0$ (charged black hole without
angular momentum).


\subsection{$J\neq 0$, $Q=0$}


The solution of the equations is
\begin{eqnarray}
f^{2} &=&r^{2}-M+\frac{J^{2}}{4r^{2}}  \label{fJ} \\
p &=&p(\infty )=-\frac{J}{2} \\
N &=&N(\infty ) \\
N^{\varphi } &=&-\frac{J}{2r^{2}}+N^{\varphi }(\infty) \,.  \label{NphiJ}
\end{eqnarray}
If one varies the reduced action (\ref{redhamilaction}) one picks up a
surface term \cite{Regge-Teitelboim}
\begin{equation}
(t_{2}-t_{1})2\pi [-\delta f^{2}(\infty )N(\infty )-2\delta
p(\infty)N^{\varphi }(\infty )]\,,
\end{equation}
which identifies $\delta f^{2}(\infty )=\delta M$ and $\delta
p(\infty)=-\delta J/2$. In order to have a black hole for $M\geq 0$ one
adjusts the zero of the energy so that $M$ is indeed the mass. In the same
manner, in order to have a static black hole for $J=0$ one takes $J$ as the
total angular momentum. Similarly with the electric charge, which is
identified with the integration constant $Q$ appearing in (\ref
{redhamilaction}) because its variation in the action (\ref{hamilaction})
multiplies $A_{0}(\infty )$.

The function $f^{2}$ has zeros at $r_{+}$ and $r_{-}$, the outer and inner
horizons, given by
\begin{eqnarray}
r_{+} &=&\left[ \frac{M+(M^{2}-J^{2})^{1/2}}{2}\right] ^{1/2} \,, \\
r_{-} &=&\left[ \frac{M-(M^{2}-J^{2})^{1/2}}{2}\right] ^{1/2} \,,
\end{eqnarray}
which exist provided
\begin{equation}
J^{2}\leq M^{2} \,,
\end{equation}
and coalesce if $J=M$ (extreme case).

Note that if $J\rightarrow 0$ ($M$ fixed) the root $r_{-}$ tends to $r=0$
which is considered as a (``chronological") singularity \cite{BHTZ}. Note
also that when $J=0$, $f^{2}=0$ has only one root $r=r_{+}$. Thus the limit
of the root $r_{-}$ (which is zero) is not a root of the limit of $f^{2}$.
In this sense the limit is discontinuous. This discontinuous behavior shows
itself in the graph of $f^{2}$ as a function of $r$. The minimum of $f^{2}$
occurs at $r_{\min }^{2}=(\frac{J}{2})^{2}$ whereas the value of $f^{2}$ at
the minimum is $f_{\min }^{2}=-M$ for all $J$. Thus when $J\rightarrow 0$
the whole branch of $f^{2}$ to the left of the minimum disappears. We shall
see similar behaviors for $J=0$, $Q\neq 0$ and $J\neq 0$, $Q\neq 0$.


\subsection{$J= 0$, $Q\neq 0$}

\label{solQnz}

Now one has
\begin{eqnarray}
f^{2} &=&r^{2}-M-\frac{1}{4}Q^{2}\ln r^{2}\,  \label{fQ} \\
p &=&0\, \\
N &=&N(\infty )\, \\
N^{\varphi } &=&0\,.
\end{eqnarray}
If one varies the action, one picks up a surface term
\begin{equation}
(-\delta M-\frac{1}{2}\delta Q^{2}\ln r)N(r)\,,  \label{deltafQ}
\end{equation}
which diverges if $r\rightarrow \infty $.

This divergence may be handled as follows. One encloses the system in a
large circle of radius $r_{0}$ and rewrites (\ref{fQ}) as
\begin{equation}
f^{2}=r^{2}-M(r_{0})-\frac{1}{2}Q^{2}\ln \frac{r}{r_{0}}\,,
\end{equation}
with
\begin{equation}
M=M(r_{0})-\frac{1}{2}Q^{2}\ln r_{0}\,.  \label{suma}
\end{equation}
Then (\ref{deltafQ}) is replaced by
\begin{equation}
(-\delta M(r_{0})-\frac{1}{2}\delta Q^{2}\ln \frac{r}{r_{0}})N(r)\,,
\end{equation}
and the second term vanishes when $r\rightarrow r_{0}$.

One might call $M(r_{0})$ ``the energy within the radius $r_{0}$''. It
differs from $M$ by $-\frac{1}{2}Q^{2}\ln r_{0}$ which may be thought of as
the electrostatic energy outside $r_{0}$ up to an (infinite) constant which
is absorbed in $M(r_{0})$. The sum (\ref{suma}) is then independent of $%
r_{0} $, finite and equal to the total mass.

Thus in practice one does not bring in $r_{0}$ and writes
\begin{equation}
`` \lim_{r\rightarrow \infty }\delta (Q^{2}\ln r)=0 \, " \,.
\end{equation}

A similar situation occurs in four spacetime dimensions. One may write the
Reissner-Nordstr\"{o}m $g_{00}$ as $1-\frac{2M(r_{0})}{r}+Q^{2}\left( \frac{1%
}{r}-\frac{1}{r_{0}}\right) $. Then $M=M(r_{0})+ \frac{Q^{2}}{2r_{0}}$ is
the total mass, independent of $r_{0}$, and $\frac{Q^{2}}{2r_{0}}$ is the
electrostatic energy outside a sphere of radius $r_{0}$. In this case it is
not necessary to include an infinite constant in $M(r_{0})$ since $\frac{%
Q^{2}}{2r_{0}}$ vanishes when $r_{0} \rightarrow \infty $.

The function $f^{2}$ given by (\ref{fQ}) tends to positive infinity if $%
r\rightarrow 0$ or if $r\rightarrow \infty $. It has a minimum at $r=|Q|/2$.
The value of $f^{2}$ at the minimum is
\begin{equation}
-M+\left( \frac{Q}{2}\right) ^{2}\left[ 1-\ln \left( \frac{Q}{2}\right)
^{2}\right] \,.  \label{min}
\end{equation}
If (\ref{min}) is negative there are two zeros of $f^{2}$, $r_{+}$ and $%
r_{-} $. If (\ref{min}) vanishes the two roots coincide and one has an
extreme black hole. The situation is illustrated in Fig. 1.

\vskip 0.1cm
\begin{center}
\epsfxsize=5cm
\leavevmode
\epsfbox{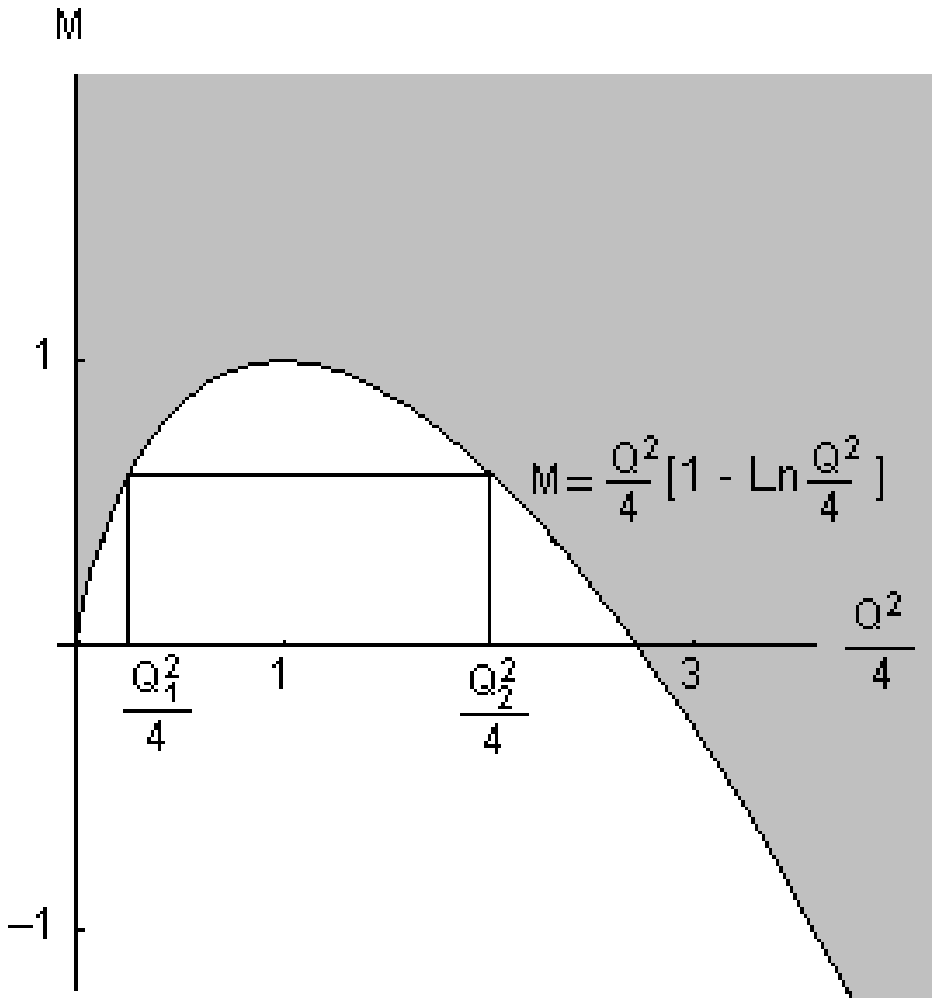}
\end{center}
{\small {\bf Figure 1}: \medskip Region in the mass-charge plane
for which there are black holes. Black holes exist whenever
$M-\left( \frac{Q}{2}\right) ^{2}\left[ 1-\ln \left(
\frac{Q}{2}\right) ^{2}\right] \geq 0.$ This corresponds to the
shaded area in the diagram. Extremal black holes are at the
boundary between the shaded and unshaded areas. Note that if the
electric charge is large enough black hole solutions exist for
arbitrarily negative values of the mass.}

\vskip 0.5cm

Now, the function $\left( \frac{Q}{2}\right) ^{2}\left[ 1-\ln \left( \frac{Q%
}{2}\right) ^{2}\right] $ vanishes at $Q=0$, has a maximum at $\left( \frac{Q%
}{2}\right) ^{2}=1$ with value 1, vanishes at $\left( \frac{Q}{2}\right)
^{2}=e$ and tends to negative infinity for large $\left( \frac{Q}{2}\right)
^{2}$. This means that if $M>1$ there are always two roots $r_{\pm }$ which
are different. When $M=1$ the two roots coincide for $\left( \frac{Q}{2}%
\right) ^{2}=1$ . If $0<M<1$ there are two branches $Q^{2}\leq Q_{{\rm 1}}^{2}$
and $Q^{2}\geq Q_{{\rm 2}}^{2}$, where $Q_{{\rm 1}}^{2}<Q_{{\rm 2}}^{2}$ are
the two roots of (\ref{min}).

One sees that if the electric charge is large enough black hole solutions
exist even for negative values of the mass. This feature is in sharp
contrast with what happens in $3+1$ dimensions and makes the electrically
charged $2+1$ black hole somewhat pathological.

Lastly, consider the limit $Q\rightarrow 0$. The situation is analogous to
the limit $J\rightarrow 0$ discussed at the end of Sec. (3.1). Now one has $%
r_{\min }^{2}=\left( \frac{Q}{2}\right) ^{2}$ and $f_{\min }^{2}\rightarrow
-M$ so that, again, the whole ascending branch to the left of the minimum
disappears in the limit $Q\rightarrow 0$. Note that $r_{-}$ approaches zero
as $r_{-}\sim e^{-2M/Q^{2}}$.


\subsection{$M\neq 0$, $J\neq 0$, $Q\neq 0$ small}

\label{solpert}

One can solve equations (\ref{one})--(\ref{te1}) perturbatively for small $Q$%
. This is of interest because the solution for small $Q$ captures the
behavior at large distances for a generic $Q$, and also because it gives
insight into the properties of horizons.

We start with expressions (\ref{fJ})--(\ref{NphiJ}) which we will denote
with a subscript zero (``unperturbed'') and add to them corrections of order
$Q^{2}$. One has ${\cal B}_{0}=0$. The equations for the perturbation then
read
\begin{eqnarray}
(\Delta f^{2})^{\prime } &=&\frac{2J}{r^{3}}\Delta p-\frac{Q^{2}}{2r}-\frac{%
f_{0}^{2}\Delta {\cal B}^{2}}{2r}  \label{1} \\
\Delta p^{\prime } &=&-\frac{Q\Delta {\cal B}}{2}  \label{2} \\
\Delta N^{\prime } &=&\frac{\Delta {\cal B}^{2}}{2r}  \label{3} \\
(\Delta N^{\varphi })^{\prime } &=&\frac{J}{r^{3}}\Delta N-\frac{2}{r^{3}}%
\Delta p  \label{4} \\
\left( \frac{(r^{2}-r_{+}^{2})(r^{2}-r_{-}^{2})}{r^{3}}\Delta {\cal B}%
\right. &+&\left. \frac{JQ}{2r^{2}}-N^{\varphi }(\infty )Q\right) ^{\prime
}=0\,.  \label{5}
\end{eqnarray}

These equations must be integrated demanding that $N(\infty )$, $%
N^{\varphi}(\infty )$, $M$ and $J$ be unchanged. Thus we should ask
\[
\Delta N(\infty)=0 \,, \Delta N^{\varphi }(\infty )=0\,.
\]
and also demand that $\Delta p(r)$ and $\Delta f^{2}$ should
vanish as $r \rightarrow \infty $, up to logarithmic terms
multiplied by $Q^{2}$, which are not considered when identifying
$J$ and $M$ by the reasoning of Section 3.2. The integration
constant in (\ref{5}) is fixed by a regularity requirement as
discussed below.

One obtains
\begin{eqnarray}
\Delta f^{2} &=&-\frac{Q^{2}}{4}\left( 1+\frac{r_{-}^{2}}{r_{+}^{2}}-2\frac{%
r_{-}^{2}}{r^{2}}\right) \ln (r^{2}-r_{-}^{2})  \label{11} \\
\Delta p &=&-\frac{JQ^{2}}{8r_{+}^{2}}\ln (r^{2}-r_{-}^{2})  \label{21} \\
\Delta N &=&-\frac{Q^{2}r_{-}^{2}}{4r_{+}^{2}}\frac{1}{r^{2}-r_{-}^{2}}
\label{31} \\
\Delta N^{\varphi } &=&-\frac{JQ^{2}}{8r_{+}^{2}r^{2}}\left( 1+\ln
(r^{2}-r_{-}^{2})\right)  \label{41} \\
\Delta {\cal B} &=&\frac{JQ}{2r_{+}^{2}}\frac{r}{r^{2}-r_{-}^{2}}\,.
\label{51}
\end{eqnarray}
Here $r_{+}$ and $r_{-}$ are those of the unperturbed solution. Note that
when $J=0$ ($r_{-}=0$) the results go over into the exact expressions found
in the previous subsection.

The integration constant in (\ref{5}) has been chosen so that the
magnetic field referred to an orthonormal basis,
\begin{equation}
\hat{B}=F_{\hat{r}\hat{\varphi}}=g^{-1/2}{\cal B}=\frac{f{\cal B}}{r},
\end{equation}
is regular at the (outer) horizon $r_+$. Indeed, (\ref{51}) is equivalent to
writing
\begin{equation}
{\cal B}(r)=\frac{Q[N^{\varphi }(r)-N^{\varphi }(r_+)]}{N(r)f^2(r)}r\,.
\label{te4}
\end{equation}
Note that, as a consequence, $\hat{ B}(r_+)=0$.

As a result of the perturbation the horizon is now at $r_{+}+\Delta r_{+}$
where $\Delta r_{+}$ may be obtained from
\begin{eqnarray}
\Delta r_{+}^{2} &=&-\left. \frac{\Delta f^{2}}{(df_{0}^{2}/dr_{+}^{2})}%
\right| _{r_{+}}  \nonumber \\
&=&-\frac{Q^{2}}{4}\ln [r_{+}^{2}-r_{-}^{2}]  \nonumber \\
&=&-\frac{Q^{2}}{2}\ln [M^{2}-J^{2}] \,.
\end{eqnarray}

The solution given by (\ref{11}-\ref{51}) is singular at $r_{-}$. This means
that the inner horizon of the rotating uncharged black hole is unstable
under the addition of a small electric charge.

\section{Exact solution with $M\neq 0$, $J\neq 0$, $Q\neq 0$}

The system of ordinary, coupled, non-linear, first order differential
equations (\ref{one}-\ref{te1}) appears not to be tractable directly.
However one may find the solution by means of an ``illegitimate coordinate
transformation''. In this case ``illegitimate'' means a transformation that
does not preserve the periodicity of the angular variable and also changes
the range of the radial variable. As a consequence of the change in
periodicity, the Casimir invariants of the symmetry group at infinity are
changed.

Generating new solutions from old ones by means of coordinate
transformations which are illegitimate in some intermediate step, but lead
to a sensible answer, is a procedure that has been quite useful in general
relativity although there appears to be no general rationale behind it.

For a conical singularity in three spacetime dimensions, the idea of
bringing in angular momentum through an illegitimate coordinate
transformation was used in Rfes. \cite{DJT,Brown,DM}. In \cite{Clement} the
procedure was used to generate the uncharged rotating black hole metric
(with a transformation not quite the same as the one used here), but was not
carried out properly for the charged case.

\subsection{Rotating solution with $Q=0$ revisited}

In this case one knows that the black holes with $J=0$ and $J\neq 0$ are
both obtained by identifications of anti-de Sitter space as discussed in
\cite{BHTZ}. Therefore, if one forgets the identifications there exists a
coordinate transformation that takes the standard form of the black hole
line element with $J=0$ onto the one with $J\neq 0$. It ios given by
\begin{eqnarray}
\tilde{t} &=&\frac{R_{+}}{r_{+}}(t-\frac{R_{-}}{R_{+}}\varphi )
\label{lorentz1} \\
\tilde{\varphi} &=&\frac{R_{+}}{r_{+}}(\varphi -\frac{R_{-}}{R_{+}}t)
\label{lorentz2} \\
r^{2} &=&R^{2}-R_{-}^{2}.  \label{lorentz3}
\end{eqnarray} 

This transformation may be obtained by composing the transformations that relate each of these black holes to the standard form of the anti-de Sitter line element and which are given by Eq. (3.29) of \cite{BHTZ}. The individual transformations and also their composition change the functional form of the anti-de Sitter line element and therefore do not belong to the $SO(2,2)$ symmetry group of anti-de Sitter space.

Here we have denoted by $(r,\widetilde{\varphi },\widetilde{t})$ and $%
(R,\varphi ,t)$ the standard coordinates for the non rotating and rotating
cases respectively. To obtain the rotating black hole one must demand that $%
\varphi $ be periodic with period $2\pi $ and also that $0\leq R^{2}<\infty $%
. As already anticipated, these requirements are incompatible with $%
\widetilde{\varphi }$ having period $2\pi $ and $0\leq r^{2}<\infty $. It is
important to realize that the outer horizon is the image $R_{+}$ of the
unique horizon that the non rotating black hole has at $r_{+}^{2}=\widetilde{%
M}$. On the other hand, the inner horizon comes in because the Jacobian
factor $dR/dr$ vanishes at $r=0$, whose image is $R_{-}$.

Now, the transformation (\ref{lorentz1},\ref{lorentz2},\ref{lorentz3}) has
more freedom than what we need because it takes a non-rotating black hole
with a given mass onto the most general black hole. It suffices for our
needs to be able to endow a given black hole with angular momentum, much as
one endows a particle with linear momentum by means of a boost. This analogy
is quite appropriate, indeed if we write
\begin{eqnarray}
\left( \frac{R_{-}}{R_{+}}\right) ^{2} &=&\omega ^{2}\leq 1  \label{r+/-} \\
\frac{R_{+}}{r_{+}} &=&\alpha \frac{1}{\sqrt{1-\omega ^{2}}},  \label{R+/r+}
\end{eqnarray}
we see that (\ref{lorentz1},\ref{lorentz2}) is the product of a Lorentz
boost and a conformal rescaling of $\varphi $, $t$. One may set

\begin{equation}
\alpha =1  \label{alpha}
\end{equation}
by rescaling the radial variable $R$ and the parameters $R_{-}$ and $R_{+}$
by factor $\alpha $, and we shall do so. Therefore we will take as our
``rotation boost'' the transformation
\begin{eqnarray}
\tilde{t}&=&\frac{t-\omega \varphi }{\sqrt{1-\omega ^{2}}}, \\
\widetilde{\varphi }&=&\frac{\varphi -\omega t}{\sqrt{1-\omega ^{2}}},
\label{boost}
\end{eqnarray}
supplemented by the change of the radial coordinate
\begin{equation}
r^{2}=R^{2}-\frac{\omega ^{2}r_{+}^{2}}{1-\omega ^{2}}.  \label{R(r)}
\end{equation}

The reason for adopting this particular way of eliminating one of the
redundant parameters in (\ref{lorentz1},\ref{lorentz2}) (rather than, say,
setting $R_{+}/r_{+}=1$) is, basically, that it simplifies considerably the
analysis for the charged case. On less pragmatic grounds, taking (\ref
{lorentz1},\ref{lorentz2}) to be just a Lorentz boost is appealing because
it captures distinctly the limiting nature of the extreme black hole which
appears as the analogue of a zero mass particle.

Indeed, one may rewrite (\ref{r+/-}) and (\ref{R+/r+}) as

\begin{eqnarray*}
R_{+} &=&\frac{r_{+}}{\sqrt{(1-\omega ^{2})}} \\
R_{-} &=&\frac{r_{+}\omega }{\sqrt{(1-\omega ^{2})}}.
\end{eqnarray*}
Thus we see that $(R_{+,}R_{-})$ transforms as a vector under the boost and $%
r_{+}=\widetilde{M}^{1/2}$ plays the role of the rest mass. The extreme
black hole corresponds to the limit $\omega ^{2}\rightarrow 1$, $\widetilde{M%
}\rightarrow 0$, keeping $R_{+}^{2}=\widetilde{M}/(1-\omega ^{2})$ finite.

The corresponding transformation formulas for the mass and the angular
momentum read

\begin{equation}
M=\widetilde{M}(1+\omega ^{2})/(1-\omega ^{2}),  \label{M/Q=0}
\end{equation}
\begin{equation}
J=2\omega \widetilde{M}/(1-\omega ^{2}).  \label{J/Q=0}
\end{equation}

Recall that $M$ and $J$ are precisely the Casimir invariants associated with the identification which produces the black hole out of anti-de Sitter space \cite{BHTZ}. It follows from (\ref{M/Q=0}) and (\ref{J/Q=0}) that each of these Casimir invariants separately changes under the boost; however, the combination $M^{2}-J^{2}=r_{+}^{4}$ remains unchanged.

\subsection{Charged case}

When $Q\neq 0$, the black hole is not obtained through identifications of points in anti-de Sitter space or in another universal space. Therefore the reasoning acompanying equations (\ref{lorentz1})-(\ref{lorentz3}) no longer applies. Nevertheless, we will again bring in angular momentum through the rotation boost (\ref{boost}) supplemented by a change of the radial coordinate which will be a generalization of (\ref{R(r)}) and which we will derive below. The procedure will of course produce a solution of the field equations since one is performing a change of coordinates. That the solution is a charged rotating black hole will follow from verifying that the asymptotic conditions (\ref{11}-\ref{51}) are met and because it will have the appropriate horizon structure. 

We start from a line element of the form,

\begin{equation}
ds^{2}=-f^{2}(r)d\tilde{t}^{2}+f^{-2}(r)dr^{2}+r^{2}d\tilde{\varphi}^{2},
\end{equation}
and apply the boost (\ref{boost}) to it  to obtain
\begin{equation}
ds^{2}=-N^{2}F^{2}dt^{2}+F^{-2}dR^{2}+R^{2}(d\varphi +N^{\varphi }dt)^{2}.
\label{dsrot}
\end{equation}
with
\begin{eqnarray}
R^{2} &=&\frac{r^{2}-\omega ^{2}f^{2}}{1-\omega ^{2}}  \label{R2(f)} \\
F^{2} &=&\left( \frac{dR}{dr}\right) ^{2}f^{2}  \label{F2} \\
N &=&\frac{r}{R}\left( \frac{dr}{dR}\right) =\left( \frac{dr^{2}}{dR^{2}}%
\right)  \label{N} \\
N^{\varphi } &=&\frac{\omega (f^{2}-r^{2})}{(1-\omega ^{2})R^{2}}.
\label{Nphi}
\end{eqnarray}

To boost to rotation the uncharged black hole one substitutes $%
r^{2}-r_{+}^{2}$ for $f^{2}$ in (\ref{R2(f)}-\ref{Nphi}) and obtains the
uncharged rotating black hole metric, with (\ref{R2(f)}) giving the
transformation (\ref{R(r)}) of the radial coordinate. If one starts instead
with the charged nonrotating black hole, one takes

\begin{equation}
f^{2}=r^{2}-\widetilde{M}-\frac{1}{4}{\widetilde{Q}}^{2}\ln r^{2}\,.
\end{equation}
and obtains, after the boost,

\begin{eqnarray}
R^{2} &=&r^{2}+\frac{\omega ^{2}}{(1-\omega ^{2})}(\widetilde{M}+\frac{{%
\widetilde{Q}}^{2}}{4}\ln r^{2})  \label{R2Q} \\
F^{2} &=&\frac{\left( r^{2}+\frac{\omega ^{2}{\widetilde{Q}}^{2}}{4(1-\omega
^{2})}\right) ^{2}}{R^{2}r^{2}}(r^{2}-\widetilde{M}-\frac{1}{4}{\widetilde{Q}%
}^{2}\ln r^{2}) \\
N &=&\frac{r^{2}}{r^{2}+\frac{\omega ^{2}{\widetilde{Q}}^{2}}{4(1-\omega
^{2})}} \\
N^{\varphi } &=&-\omega \frac{\widetilde{M}+\frac{1}{4}{\widetilde{Q}}%
^{2}\ln r^{2}}{(1-\omega ^{2})R^{2}}  \label{NphiQ} \\
\frac{dR}{dr} &=&\frac{r^{2}+\frac{\omega ^{2}{\widetilde{Q}}^{2}}{%
4(1-\omega ^{2})}}{Rr}\,.  \label{JQ}
\end{eqnarray}

The potential $A=-{\widetilde{Q}} \ln r d\tilde{t}$ transforms into
\begin{equation}  \label{AJ}
A=- \frac{\widetilde{Q}}{\sqrt{1-\omega^2}} \ln r (dt-\omega d\varphi)
\end{equation}
yielding a magnetic density
\begin{equation}
{\cal B}= \frac{\widetilde{Q} \omega R}{\sqrt{1-\omega^2}(r^2+ \frac{%
\omega^2 {\ \widetilde{Q}}^2}{4(1-\omega^2)})} \,.
\end{equation}
The momentum $p$ is determined using (\ref{four})
\begin{equation}
p=-\frac{\omega}{1-\omega^2} \left(\widetilde{M}+\frac{{\widetilde{Q}}^2}{4}
(\ln r^2 -1)\right) \,.
\end{equation}

The graph of $R^{2}$ as a function of $r^{2}$ given by (\ref{R2Q}) is shown
in Figure 2. One sees that the function has two branches separated by a
singularity in the form of an infinite throat at $r^{2}=0$. The black hole
space corresponds to the piece to the right of the throat for which $%
R^{2}\geq 0$. In the limit $Q\rightarrow 0$ the left and right sides of the
throat annihilate each other and the two branches merge on the segmented
straight line. The inner horizon of the uncharged black hole corresponds to
the intersection of the straight line and the horizontal axis, which occurs
at a negative value of $r^{2}$. When the electric charge is turned on, the
throat comes in and the old inner horizon becomes thus disconnected from the
black hole space, while a new inner horizon appears in the black hole space.
This is the origin of the perturbative instability of the inner horizon
found in Section 3. Both the
outer and inner horizons of the charged rotating black hole come from the
corresponding ones of the original charged black hole with $J=0$. They are
the images under the boost of the zeros $r_{+}$ $r_{-}$ of $f^{2}.$ Unlike
what happens in the uncharged case, the Jacobian $dR/dr$ in (\ref{JQ}) is
different from zero and does not bring in new zeros of $F^{2}$.

\vskip 1cm
\begin{center}
\epsfxsize=10cm
\leavevmode
\epsfbox{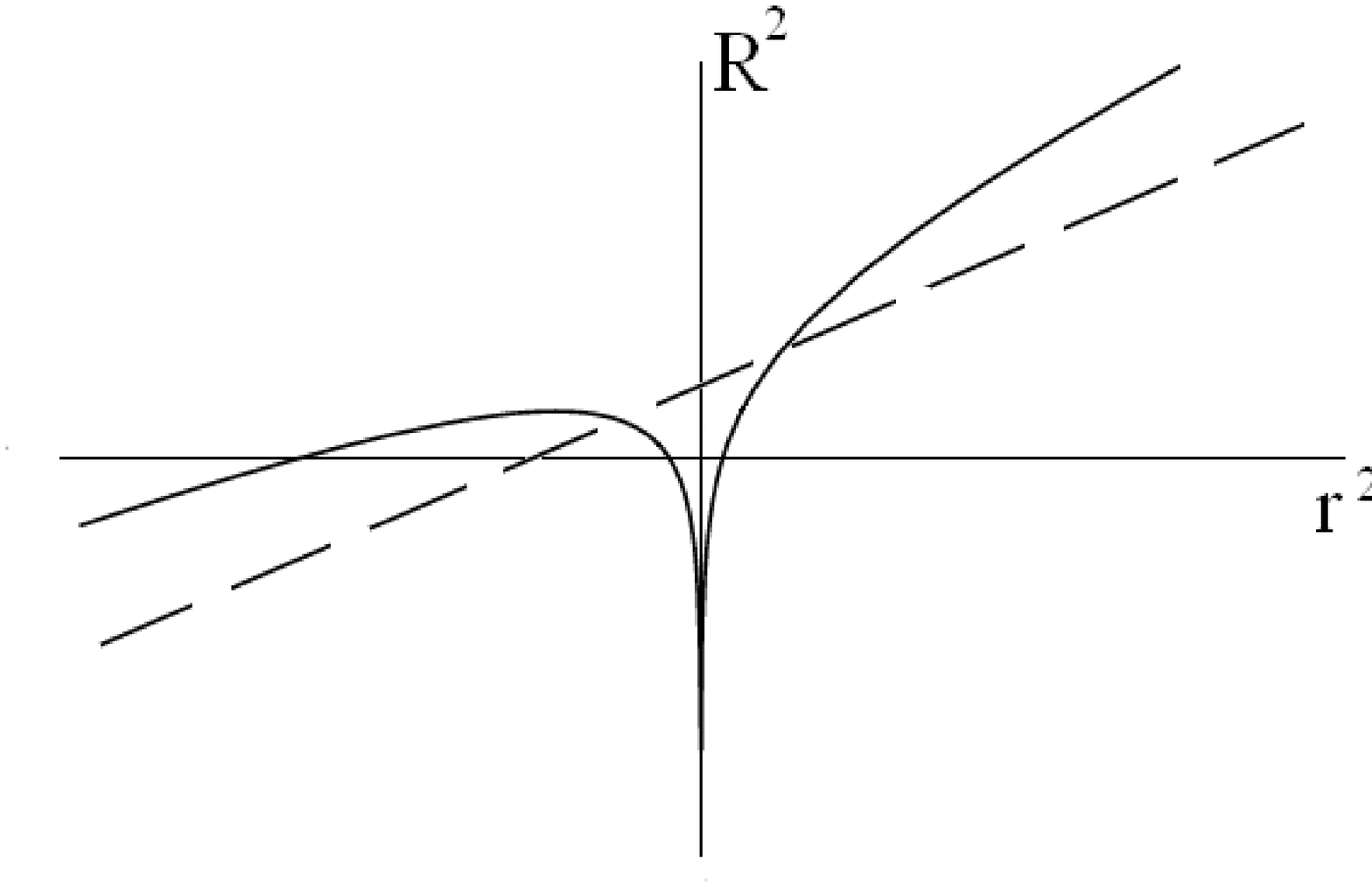}
\end{center}
\small{{\bf Figure 2:} $R^{2}$ as a function of $r^{2}$. The black hole space corresponds
to the piece to the right of the throat for which $R^{2}\geq 0$. In the
limit $Q\rightarrow 0$ the left and right sides of the throat annihilate
each other and the two branches merge on the segmented straight line.  The separation between the solid and the segmented line behaves asymptotically as $\frac{\omega ^{2}}{1-\omega ^{2}}\frac{{
\widetilde{Q}}^{2}}{4}\ln r^{2}$. The
inner horizon of the uncharged black hole corresponds to the intersection of
the straight line and the horizontal axis, which occurs at a negative value
of $r^{2}$. When the electric charge is turned on, the throat comes in and
the old inner horizon becomes thus disconnected from the black hole space,
while a new inner horizon appears in the black hole space. This is the
origin of the perturbative instability of the inner horizon.}
\vskip 0.5cm

\subsection{ Asymptotic form}

We now verify that the rotating solution has the correct asymptotic form and
relate the parameters $\widetilde{M}$, $\widetilde{Q}$ to $M$, $Q$ and $J$,
the actual mass, charge and angular momentum of the charged rotating
solution.

For $R\rightarrow \infty $, Eq. (\ref{R2Q}) takes the form
\begin{equation}
r^{2}=R^{2}-\frac{\omega ^{2}}{1-\omega ^{2}}\left( \widetilde{M}+\frac{{%
\widetilde{Q}}^{2}}{4}\ln R^{2}\right) +O(1/R^{2})\,.
\end{equation}
Using this relation one obtains the asymptotic forms of $F^{2}$ and $p$
\begin{eqnarray}
F^{2} &=&R^{2}-\frac{1}{1-\omega ^{2}}\left( \widetilde{M}(1+\omega ^{2})-%
\frac{\omega ^{2}{\widetilde{Q}}^{2}}{2}\right)  \nonumber  \label{F2asym} \\
&-&\frac{(1+\omega ^{2}){\widetilde{Q}}^{2}\ln R^{2}}{4(1-\omega ^{2})}%
+O(1/R^{2}), \\
p &=&-\frac{\omega }{1-\omega ^{2}}\left( \widetilde{M}-\frac{{\widetilde{Q}}%
^{2}}{4}\right) -\frac{\omega {\widetilde{Q}}^{2}}{1-\omega ^{2}}\ln
R^{2}+O(1/R^{2}),  \label{pasym}
\end{eqnarray}
whereas from (\ref{AJ}) one has
\begin{equation}
{\cal E}^{R}=\frac{\widetilde{Q}}{\sqrt{1-\omega ^{2}}}.  \label{ER}
\end{equation}
From (\ref{F2asym}), (\ref{pasym}) and (\ref{ER}) one finds that the
symptotic forms are the desired ones provided that
\begin{eqnarray}
M &=&\frac{1}{1-\omega ^{2}}\left( \widetilde{M}(1+\omega ^{2})-\frac{\omega
^{2}{\widetilde{Q}}^{2}}{2}\right) \,,  \label{newM} \\
J &=&\frac{2\omega }{1-\omega ^{2}}\left( \widetilde{M}-\frac{{\widetilde{Q}}%
^{2}}{4}\right),  \label{newJ}
\end{eqnarray}
and
\begin{equation}
Q=\frac{\widetilde{Q}}{\sqrt{1-\omega ^{2}}}\,.  \label{newQ}
\end{equation}
It may seem surprising that $Q$ differs from $\widetilde{Q}$. This comes
from the change in the periodicity of $\varphi $ under the boost, which
changes the region of integration of the charge density.

Now, for a black hole one wants to express the metric in terms of the
charges at infinity (hairs). To this end, one should solve Eqs. (\ref{newM})(%
\ref{newJ})(\ref{newQ}) to express $\widetilde{M}$, $\widetilde{Q}$ and $%
\omega $ as functions of $M$, $Q$,and $J$. If one can do this, the ``rest
mass'' $\widetilde{M}(M,Q,J)$ and ``rest charge'' $\widetilde{Q}(M,Q,J)$ are
by construction invariants under the boost, and generalize the $\widetilde{M}
$ of the case without charge, whereas $\omega (M,Q,J)$ is the ``velocity
expressed in terms of the momentum''.

The difficult part in inverting Eqs. (\ref{newM})(\ref{newJ})(\ref{newQ}) is
to solve the cubic equation that they imply, namely,
\begin{equation}
\frac{Q^{2}}{4}\omega ^{3}-\frac{J\omega ^{2}}{2}+\omega \left( M-\frac{Q^{2}%
}{4}\right) -\frac{J}{2}=0\,.  \label{cubic}
\end{equation}
To gain insight into this equation it is useful to consider the case $Q=0$
which may be rewritten as
\begin{equation}
\omega ^{2}-\frac{2M}{J}\omega +1=0.  \label{Q=0}
\end{equation}
The solution of this equation, which satisfies $\omega ^{2}<1$, is

\begin{equation}\label{SolQ=0}
\omega =\mbox{\rm{sgn}}\left( \frac{M}{J}\right) \left( \left| \frac{M}{J}\right| -\sqrt{%
\left( \frac{M}{J}\right) ^{2}-1}\right) ,
\end{equation}
and exists if and only if
\begin{equation}
M^{2}>J^{2}.  \label{M>J}
\end{equation}
In the limiting case $M^{2}=J^{2}$, one has $\omega \rightarrow \pm 1$.

If the solution (\ref{SolQ=0}) is inserted back in the original equations
Eqs. (\ref{M/Q=0}), (\ref{J/Q=0}), one finds
\begin{equation}
\widetilde{M}=\mbox{\rm{sgn}}M\sqrt{M^{2}-J^{2}}.
\end{equation}
We see, therefore, two points of interest. They are: ({\bf i}) A real $%
\omega $, with $\omega ^{2}<1$, exists only for a range of the parameters of
the algebraic equation (\ref{Q=0}), namely $M^{2}>J^{2}$. ({\bf ii}) Even
when $\omega ^{2}<1$ exists, the solution is not necessarily a black hole.
One needs, in addition to (\ref{M>J}), $\widetilde{M}\geq 0$, which in this
case is equivalent to $M>0$.

When $Q\neq 0$, one expects to face the same situation. There will be a
certain region in the space of parameters $M$, $J$ and $Q$ (actually, only
the two independent combinations $\frac{M}{J}$ and $\frac{Q^{2}}{4J}$ enter)
for which $\omega ^{2}<1$ exists. Within that region one will be able to
express $\widetilde{M}$ and $\widetilde{Q}$ as functions of $M$, $J$ and $Q$
and the solution will be a black hole when $\widetilde{M}$ and $\widetilde{Q}
$ obey the additional restrictions described in Figure 1. It does not seem
to be possible to exhibit explicitly the necessary and sufficient condition
for $M$, $J$ and $Q$ to be such that the cubic equation (\ref{cubic}) admits
a unique real solution $\omega ^{2}<1$. The best we have been able to do is
to show by analyzing the graph of the cubic, that a sufficient, but not
necessary conditions are $M^{2}>J^{2}$ and $M-\frac{Q^{2}}{4}>0$. It is also
straightforward to show that under these conditions $J$ and $\omega $ have
the same sign.

\noindent {\bf Acknowledgments}: We thank Prof. Marc Henneaux for
useful discussions and enlightening comments. This work was
partially supported by the grants Nos. 1970151, 1990189, 3970004
from FONDECYT (Chile). The institutional support to the Centro de
Estudios Cient\'{\i}ficos de Santiago of Fuerza A\'{e}rea de
Chile, I. Municipalidad de Las Condes, I. Municipalidad de
Santiago and a group of Chilean companies (AFP Provida, Codelco,
Copec, Empresas CMPC, Gener S.A., Minera Collahuasi, Minera
Escondida, Novagas, Business Design Associates, Xerox Chile) is
also recognized.



\begin{thebibliography}{99}
\bibitem{BTZ}  M.~Ba{\~{n}}ados, C.~Teitelboim and J.~Zanelli, Phys. Rev.
Lett. {\bf 69}, 1849 (1992).

\bibitem{BHTZ}  M.~Ba{\~{n}}ados, M.~Henneaux, C.~Teitelboim and J.~Zanelli,
Phys. Rev. {\bf D48}, 1506 (1993).

\bibitem{Carlip}  S.~Carlip, Class. Quant. Grav. {\bf 12}, 2853 (1995).

\bibitem{Others}  See for example: M. Kamara and T. Koikawa, Phys. Lett.
{\bf B 353}, 196, (1995); S. Fernando and F. Mansouri, ``Rotating
charged solutions to Einstein Maxwell Chern-Simons theory in
(2+1)-dimensions'', gr-qc/9705016; M. Cataldo and P. Salgado,
Phys. Lett.{\bf \ B448} (1999), 20; A. Garc\'{\i}a, ``On the Rotating
Charged BTZ Metric'', hep-th/9909111:

\bibitem{Coussaert-Henneaux}   O. Coussaert and M. Henneaux, Phys. Rev.
Lett. {\bf 72} (1994) 183.

\bibitem{DJT}  S. Deser, R. Jackiw and G. 't Hooft, Ann. Phys. {\bf 152},
220 (1984).

\bibitem{Brown}  J. D. Brown, {\it ``Lower dimensional gravity''}, World
Scientific Publishing, Singapore (1988).

\bibitem{DM}  S. Deser and P. O. Mazur, Class. Quantum Grav. {\bf 2}, L51
(1985).

\bibitem{Clement}  G. Cl\'{e}ment, Phys. Lett. {\bf B367}: 70-74, (1996).

\bibitem{Regge-Teitelboim}  T.~Regge and C.~Teitelboim, Ann. Phys. {\bf 88},
286 (1974).
\end{thebibliography}
\end{document}